\documentclass[12pt]{iopart}

\usepackage[english]{babel}
\usepackage[dvips]{graphicx}
\usepackage{amstext,amssymb}
\usepackage[comma,square,numbers,sort]{natbib}
\usepackage{array}
\usepackage{color}

\newcommand{\eqref}[1]{(\ref{#1})}

\newcommand{\eVq}  {\text{eV}^2}

\begin{document}


\title{Global neutrino data and recent reactor fluxes:\\ 
status of three-flavour oscillation parameters}

\author{Thomas Schwetz\dag, Mariam
  T{\'o}rtola\S\ and J. W.~F.~Valle\S}

\address{\dag\ Max-Planck-Institut f\"ur Kernphysik, PO Box 103980, 
69029 Heidelberg, Germany}

\address{\S\ AHEP Group, Instituto de F\'{\i}sica Corpuscular --
  C.S.I.C./Universitat de Val{\`e}ncia, \\
  Edificio Institutos de Paterna, Apt 22085, E--46071 Valencia, Spain}

\ead{schwetz@mpi-hd.mpg.de, mariam@ific.uv.es, valle@ific.uv.es}

\begin{abstract}
  We present the results of a global neutrino oscillation data
  analysis within the three-flavour framework. We include latest
  results from the MINOS long-baseline experiment (including electron
  neutrino appearance as well as anti-neutrino data), updating all
  relevant solar (SK II+III), atmospheric (SK I+II+III) and reactor
  (KamLAND) data.  Furthermore, we include a recent re-calculation of
  the anti-neutrino fluxes emitted from nuclear reactors. These
  results have important consequences for the analysis of reactor
  experiments and in particular for the status of the mixing angle
  $\theta_{13}$. In our recommended default analysis we find from the
  global fit that the hint for non-zero $\theta_{13}$ remains weak, at
  1.8$\sigma$ for both neutrino mass hierarchy schemes. However, we
  discuss in detail the dependence of these results on assumptions
  concerning the reactor neutrino analysis.

\vskip 1cm
\noindent
keywords: Neutrino mass and mixing; neutrino oscillation; solar and atmospheric neutrinos;
  reactor and accelerator neutrinos

\end{abstract}

\section{Introduction}

The discovery of neutrino mixing and oscillations provides firm evidence for
physics beyond Standard Model, opening a new era in particle physics.  Here
we update the three-neutrino oscillation results of 
Ref.~\cite{Schwetz:2008er}~\footnote{When referring to this paper we include
also its updates available at arXiv:0808.2016v3 [hep-ph]. Further technical
details of the analysis as well as earlier experimental references are also
given in our previous review in \cite{Maltoni:2004ei}. For other global
analyses see \cite{GonzalezGarcia:2010er, Fogli:2010zz}.} with a special
emphasis on the new reactor anti-neutrino flux results of
Refs.~\cite{Mueller:2011nm,Mention:2011rk} which have an important impact on
the determination of the mixing angle $\theta_{13}$. We include new data
from the MINOS Collaboration, both for $\nu_\mu\to\nu_e$ transitions
\cite{Adamson:2010uj} and $\nu_\mu$ disappearance \cite{vahle,
minos:2011ig}, the latest Super--Kamiokande (SK) solar~\cite{:2008zn,
Abe:2010hy} and atmospheric~\cite{Wendell:2010md} neutrino data, as well as
recent KamLAND reactor data~\cite{Gando:2010aa}. Our goal is to summarize
the results of the three-flavour neutrino oscillation analysis paying
attention to sub-leading three-flavor effects where they are most relevant,
as well as to the effects of the new anti-neutrino flux emitted from nuclear
reactors which has been re-evaluated in Ref.~\cite{Mueller:2011nm}. The
reported value of the $\bar\nu_e$ flux is about 3\% higher than from previous
calculations. This has important consequences for the interpretation of data
from reactor experiments. We discuss the implications of the new reactor
neutrino flux for the determinations of oscillation parameters, in
particular its effect on the mixing angle $\theta_{13}$. We find that due to
the new fluxes the results depend on the inclusion of short-baseline reactor
data from distances $\lesssim 100$~m.

In Sec.~\ref{sec:atm} we present the updated analysis in the
``atmospheric sector'', discussing the Super Kamiokande I+II+III
data~\cite{Wendell:2010md} in Sec~\ref{sec:SK} and the MINOS
disappearance results taking into account the solar squared-mass
splitting and discussing the slight tension between neutrino and
anti-neutrino data in Sec.~\ref{sec:minos}. In Sec.~\ref{sec:atm-th13}
we focus on the combined SK+MINOS analysis and the determination of
$\theta_{13}$ from these data. Sec.~\ref{sec:reactor} contains the
discussion of the reactor neutrino data in the light of the new
predicted anti-neutrino fluxes, and in Sec.~\ref{sec:solar} we discuss
solar neutrino data as well as KamLAND and the other reactor
experiments. The results of the global fit are summarized in
Sec.~\ref{sec:global} including a detailed discussion of the status of
$\theta_{13}$. In particular we discuss the dependence of the
$\theta_{13}$ determination upon assumptions concerning the reactor
anti-neutrino data analysis. Conclusions follow in 
Sec.~\ref{sec:summary-conclusions}.

\section{The atmospheric sector -- SK I+II+III and MINOS}
\label{sec:atm}

\subsection{Super-Kamiokande I+II+III data}
\label{sec:SK}

We include in our analysis the full sample of atmospheric neutrino
data from all three phases of the Super-Kamiokande
experiment~\cite{Wendell:2010md}, using directly the $\chi^2$ map
provided by the Super-Kamiokande collaboration.
The atmospheric neutrino oscillation analysis is performed within the
one mass scale approximation, neglecting the effect of the solar mass
splitting, as in our previous
papers~\cite{Schwetz:2008er,Maltoni:2004ei}~\footnote{Preliminary
  results towards a full 3-flavour atmospheric neutrino analysis have
  been presented in~\cite{sk-talk}, we look forward to the
  corresponding information becoming publicly available.}.
\begin{figure}
  \centering
  \includegraphics[height=6cm]{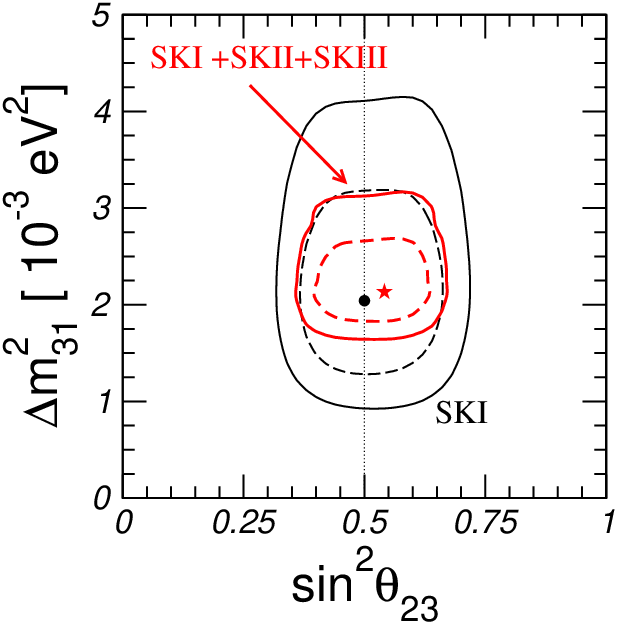}
  \caption{Comparison of our previous SK-I atmospheric analysis (in black)
    and the new analysis using the latest atmospheric data from SK-I,
    SK-II, SK-II (in red) for inverted mass hierarchy.  Best fit points  
    denoted by a dot or star follow the same colour code.}
\label{fig:atm-sk}
\end{figure}

In Fig.~\ref{fig:atm-sk} we show the results of our previous
atmospheric neutrino analysis from
Refs.~\cite{Schwetz:2008er,Maltoni:2004ei} including data from the
first phase of the Super-Kamiokande experiment, compared to the new
analysis included in this update, based on the Super-Kamiokande data
from all three phases of the experiment.
The differences between the two analyzes are clearly noticeable, with
an improved determination of both oscillation parameters.  As we will
see in the following, once data from neutrino disappearance at the
MINOS long-baseline experiments are included in the analysis, the
improvement in the determination of $\Delta m^2_{31}$ due to the
recent atmospheric neutrino data is ``hidden'' by the more
constraining restrictions imposed by long-baseline data.
Nevertheless, this new atmospheric analysis will be important when
constraining the mixing angle $\theta_{23}$ and also $\theta_{13}$.

\subsection{MINOS disappearance data}
\label{sec:minos}

At the Neutrino 2010 Conference, the MINOS Collaboration has presented new
data from their searches of $\nu_\mu$ disappearance, both from the neutrino
(7.2$\times10^{20}$ p.o.t.) and the anti-neutrino (1.71$\times10^{20}$ p.o.t)
running mode~\cite{vahle}. We perform a re-analysis of the data
from~\cite{vahle} within a full three-flavour framework using the GLoBES
software~\cite{Huber:2007ji} to simulate the experiment. In addition to
matter effects, we include also the effect of $\Delta m^2_{21}$ as well as
$\theta_{13}$ and the CP-phase $\delta$ in the analysis of the disappearance
and appearance channels. Since in our analysis MINOS data are the only ones
sensitive to the phase $\delta$ we always minimize the MINOS $\chi^2$ with
respect to $\delta$.

\begin{figure}[t]
  \centering
  \includegraphics[height=6cm]{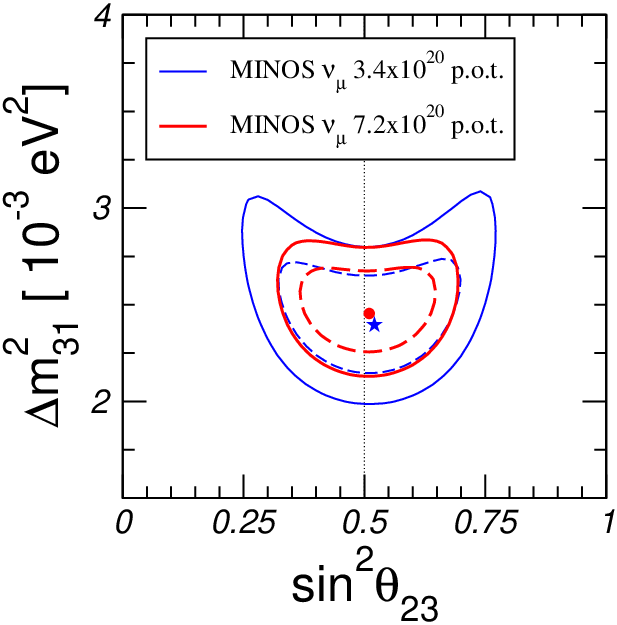}\qquad
  \includegraphics[height=6cm]{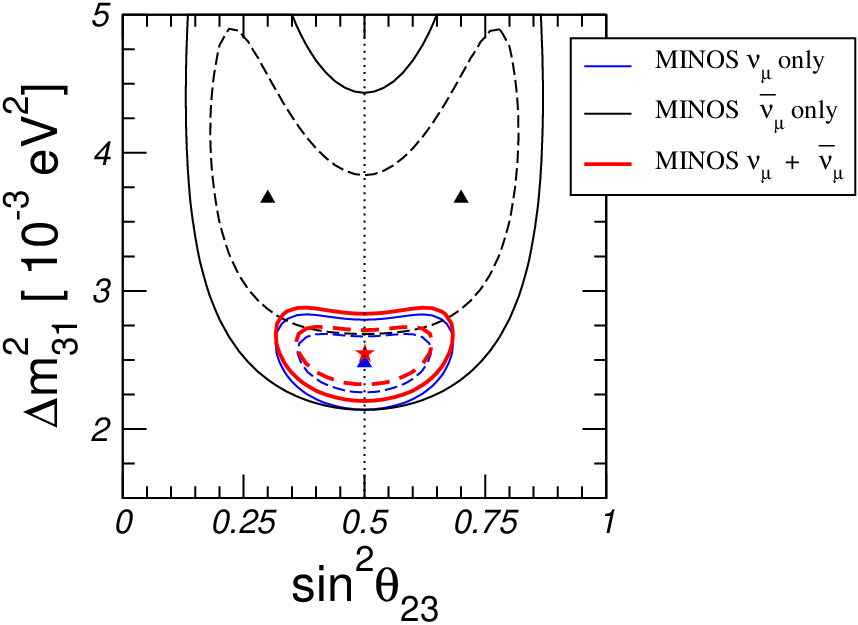}
  \caption{Determination of atmospheric neutrino parameters by the MINOS
    long-baseline experiment. Left panel: previous ($3.36 \times
    10^{20}$ p.o.t.) versus current (7.2$\times10^{20}$ p.o.t.) $\nu_\mu$ data.
    Right panel: allowed regions from recent MINOS (7.2$\times10^{20}$ p.o.t.) data, using
    neutrinos-only, anti-neutrinos-only, and the combination. Both panels
    assume normal hierarchy.  As before, best fit points follow the same colour code.
    \label{fig:min-sum2011}}
\end{figure}

In Fig.~\ref{fig:min-sum2011} we compare the analysis of new MINOS
data with the previous data release (left panel). Apart from the
smaller size of the allowed regions due to the increase in statistics,
we notice also a change in the shape of the regions. This follows from
the inclusion of the sub-leading three-flavour effects included in
this new analysis.

In the right panel we illustrate the impact of the (still small)
anti-neutrino sample. From the figure one can see that there is a
slight tension between the neutrino and anti-neutrino results: there is
no overlap of the allowed regions at less than 90\%~CL. However, at
$3\sigma$ the results of both are fully consistent. We find the
following $\chi^2$ minima and goodness-of-fit (GOF) values:
\begin{equation}
\begin{array}{rcc}
\nu:     & \chi^2_{\mathrm{min},\nu} = 24.4/(27-2)     & {\rm GOF} = 49.6\% \\
\bar\nu: & \chi^2_{\mathrm{min},\bar\nu} = 15.0/(13-2) & {\rm GOF} = 18.4\% \\
\nu+\bar\nu: & \chi^2_\mathrm{min,tot} = 46.1/(40-2)   & {\rm GOF} = 17.3\% 
\end{array}
\end{equation}
Hence the combined neutrino and anti-neutrino fit provides still an
acceptable GOF. Using the consistency test from
Ref.~\cite{Maltoni:2003cu} yields $\chi^2_\mathrm{PG} =
\chi^2_\mathrm{min,tot} - \chi^2_{\mathrm{min},\nu} -
\chi^2_{\mathrm{min},\bar\nu} = 6.6$. The value of $\chi^2_{\rm PG}$
must be evaluated for 2 degrees of freedom, which implies that
neutrino and anti-neutrino data are consistent with a probability of
3.7\%. This number indicates a slight tension between the sets, at the
level of about $2.1\sigma$. In the following we will use only neutrino
data in the global analysis. It is clear from
Fig.~\ref{fig:min-sum2011} (right) that adding also anti-neutrino data
would have negligible impact on the global result.

\subsection{The atmospheric sector: combined MINOS + atmospheric analysis and $\theta_{13}$  }
\label{sec:atm-th13}

\begin{figure}
  \centering 
  \includegraphics[height=6cm]{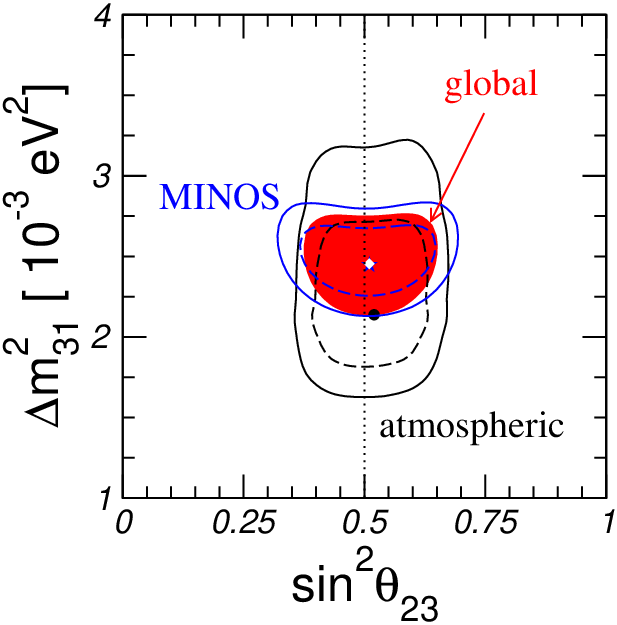}\qquad
  \includegraphics[height=6cm]{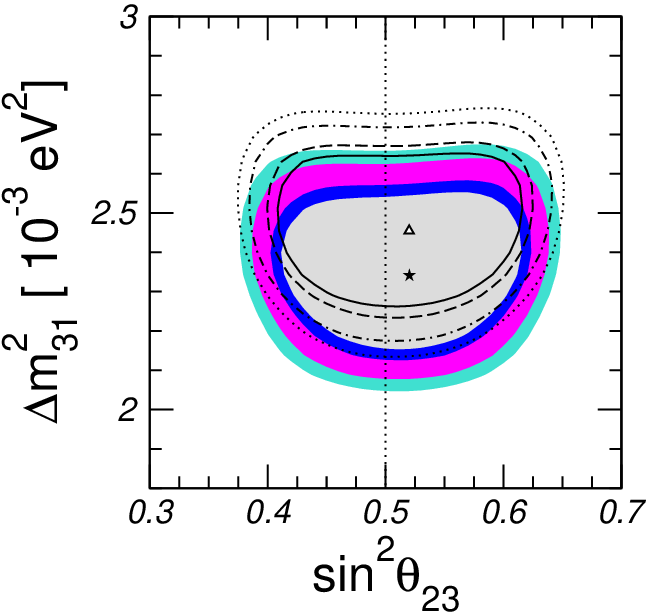} 
  \caption{Determination of the atmospheric oscillation parameters. Left:
  interplay of atmospheric (black) and MINOS disappearance (blue) data and
  the combination (red/shaded region) for normal hierarchy at 90\%~CL
  (dashed) and 3$\sigma$ (solid). Right: combined allowed regions for
  normal (black curves) and inverted hierarchy (colored regions) at 90\%,
  95\%, 99\%, 99.73\%~CL.  \label{fig:atm-sum11}}
\end{figure}

Combining the new atmospheric and MINOS disappearance data we obtain
new global constraints on the atmospheric neutrino oscillation
parameters. The results are shown in Fig.~\ref{fig:atm-sum11}.
As before, $\Delta m^2_{31}$ is determined mainly by the MINOS data,
while the atmospheric data are more important in determining the
mixing parameter $\sin^2\theta_{23}$. Notice that the sub-leading
effects of $\Delta m^2_{21}$ lead to different best fit points for
$|\Delta m^2_{31}|$, that depend on whether the mass hierarchy is
normal (NH) or inverted (IH). We find the following best fit values
with errors at $1\sigma$:
\begin{equation}\label{eq:dmq31}
|\Delta m^2_{31}| = \left\{\begin{array}{ll}
2.45\pm0.09 &\times 10^{-3} \, {\rm eV}^2 \qquad\text{(NH)} \\
2.34^{+0.10}_{-0.09} &\times 10^{-3} \, {\rm eV}^2 \qquad\text{(IH)} 
\end{array}\right.
\end{equation}
The corresponding allowed regions are shown in
Fig.~\ref{fig:atm-sum11} (right). The reason for this apparent shift
is just a result of our parameterization, using $\Delta m^2_{31}$ for
both hierarchies, and changing only the sign of it to distinguish
them. Hence, in NH the ``largest'' frequency is given by $|\Delta
m^2_{31}|$, while in IH it is $|\Delta m^2_{31}| + \Delta m^2_{21}$,
which explains why $|\Delta m^2_{31}|$ is smaller for
IH. Eq.~\ref{eq:dmq31} shows that these sub-leading effects must be
included given the present accuracy, since they are at the level of
the $1\sigma$ error on $\Delta m^2_{31}$.

Now we move to appearance MINOS data. The MINOS Collaboration has recently
also reported new data from the search of $\nu_\mu \to \nu_e$ transitions in
the Fermilab NuMI beam~\cite{Adamson:2010uj}. The new data are based on a
total exposure of 7$\times$10$^{20}$ protons-on-target, more than twice the
size of the previous data release~\cite{Adamson:2009yc}. The new MINOS far
detector data consists of 54 electron neutrino events, while, according to
the measurements in the MINOS Near Detector, $49.1 \pm 7.0 (stat) \pm 2.7
(syst)$ background events were expected. Hence the observed number of events
is in agreement with background expectations within 0.7$\sigma$ and the hint
for a non-zero value of $\theta_{13}$ present in previous
data~\cite{Adamson:2009yc} has largely disappeared. In fact, we see that
once we include the new MINOS data in our analysis, a smaller best fit point
of $\theta_{13}$ is obtained and, as a result, the hint for $\theta_{13}$ is
less significant than before: for both hierarchies we find only a
0.8$\sigma$ hint when using new MINOS data versus 1.3$\sigma$ obtained with
the previous MINOS appearance data, see e.g.~\cite{Mezzetto:2010zi} for a
discussion.

\begin{figure}
  \centering
  \includegraphics[height=6cm]{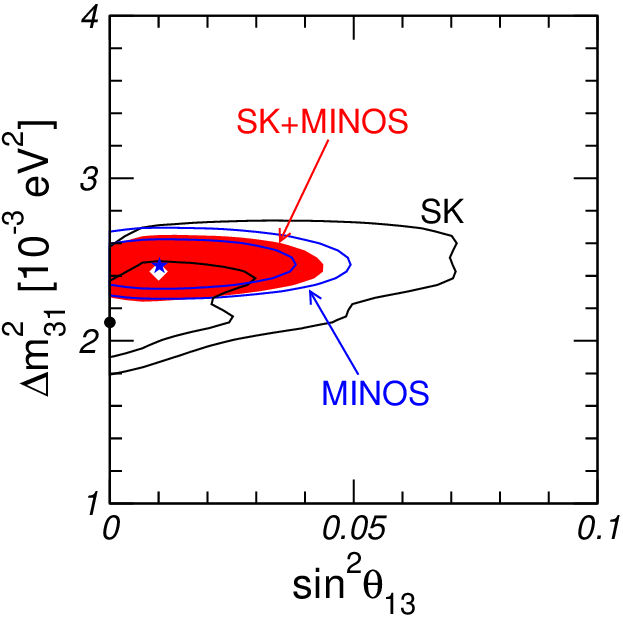}
  \includegraphics[height=6cm]{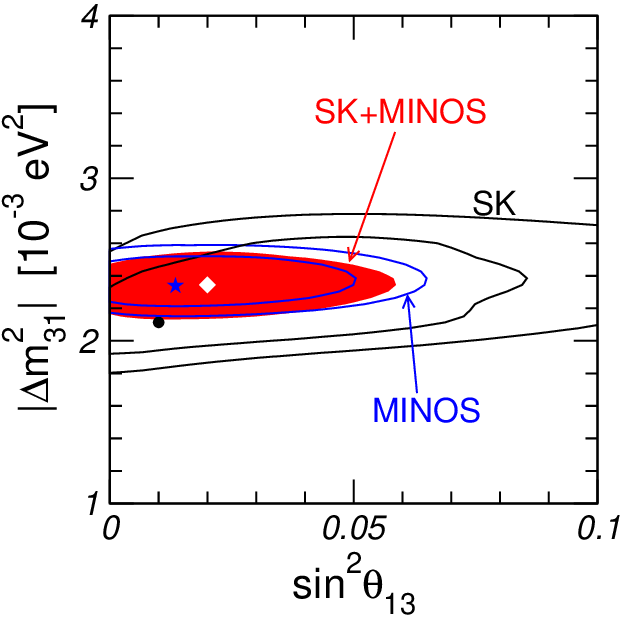}
  \caption{Allowed regions at $1\sigma$ and 90\% CL for atmospheric
    (SK) and MINOS disappearance and appearance data in the plane of
    $\sin^2\theta_{13}$ and $\Delta m^2_{31}$ for NH (left) and IH
    (right). Combined data is shown as shaded/red region at 90\% CL. The
    black dot, blue star, white diamond correspond to the best fit points of
    SK, MINOS, SK+MINOS, respectively. 
    \label{fig:th13-dmq-atm-minos}}
\end{figure}

Atmospheric neutrino data from Super-Kamiokande I+II+III described in
the previous section implies a best fit point very close to
$\theta_{13} = 0$ \cite{Wendell:2010md}, with $\Delta\chi = 0.0 (0.3)$
for $\theta_{13}=0$ for NH (IH). However, in the combination with
MINOS disappearance and appearance data we even find a slight
preference for $\theta_{13} > 0$, with $\Delta\chi^2 = 1.6 (1.9)$ at
$\theta_{13}=0$ for NH (IH). As shown in
Fig.~\ref{fig:th13-dmq-atm-minos} this happens due to a small mismatch
of the best fit values for $|\Delta m^3_{31}|$ at $\theta_{13}=0$,
which can be resolved by allowing for non-zero values of $\theta_{13}$
\cite{GonzalezGarcia:2010er}. This is similar to the hint for
$\theta_{13} > 0$ coming from a slight tension between solar and
KamLAND data, see Ref.~\cite{Schwetz:2008er}. Therefore, the hint for
$\theta_{13} > 0$ from atmospheric + LBL data becomes slightly
stronger compared to the previous data.

\section{New reactor fluxes and  implications for
oscillation parameters}
\label{sec:reactor}

Up to very recently the interpretation of neutrino oscillation
searches at nuclear power plants was based on the calculations of the
reactor $\bar\nu_e$ flux from Ref.~\cite{schreckenbach}. Indeed, the
observed rates at all reactor experiments performed so-far at
distances $L \lesssim 1$~km are consistent with these fluxes,
therefore setting limits on $\bar\nu_e$ disappearance. Recently the
flux of $\bar\nu_e$ emitted from nuclear power plants has been
re-evaluated \cite{Mueller:2011nm}, yielding roughly 3\% higher
neutrino fluxes than assumed previously. As discussed in
Ref.~\cite{Mention:2011rk} this might indicate an anomaly in reactor
experiments at $L \lesssim 1$~km, which according to the new fluxes
observe a slight deficit. For the Chooz and Palo Verde experiments at
$L \simeq 1$~km a non-zero $\theta_{13}$ could lead to $\bar\nu_e$
disappearance accounting for the reduction of the rate.  However,
$\Delta m^2_{13}$ and $\theta_{13}$ driven oscillations will have no
effect in short-baseline (SBL) experiments with $L \lesssim 100$~m.

Motivated by this situation we include here also the SBL reactor
experiments Bugey4~\cite{Declais:1994ma},
ROVNO~\cite{Kuvshinnikov:1990ry}, Bugey3~\cite{Declais:1994su},
Krasnoyarsk~\cite{kras}, ILL~\cite{Kwon:1981ua}, and
G\"osgen~\cite{Zacek:1986cu} via the rate measurements summarized in
Table~II of \cite{Mention:2011rk}, in addition to the fit of the
KamLAND, Chooz~\cite{Apollonio:2002gd} and Palo
Verde~\cite{Boehm:2001ik} experiments. We use the neutrino fluxes from
the isotopes $^{235}$U, $^{239}$Pu, $^{238}$U, $^{241}$Pu obtained in
\cite{Mueller:2011nm}. For each reactor experiment we take into
account the appropriate relative contribution of the isotopes to the
total flux and we include the uncertainty on the integrated flux for
each isotope given in Table~I of \cite{Mention:2011rk}, correlated
among all experiments. The total error on fluxes from
$^{235}$U, $^{239}$Pu, $^{241}$Pu are at the level of 2\%, where we
assume that an error of 1.8\% is fully correlated among the three
isotopes, due to a common normalization uncertainty of the
corresponding beta-spectra measured in \cite{schreckenbach}.

\begin{figure}
  \centering
  \includegraphics[height=6cm]{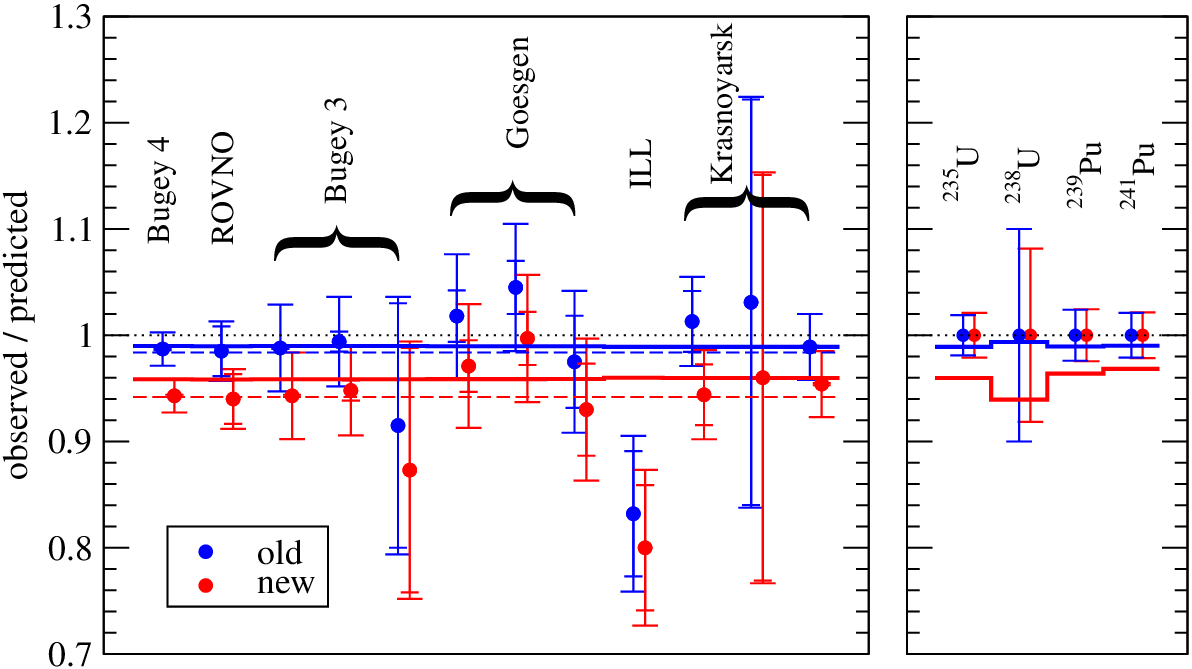}
  \caption{Short-baseline reactor data. We show the observed rate
  relative to the predicted rate based on old \cite{schreckenbach} (blue)
  and new \cite{Mueller:2011nm, Mention:2011rk} (red) flux calculations.
  Small error bars show statistical and uncorrelated systematic uncertainties, large error
  bars include in addition correlated systematic Al uncertainties. The solid
  histograms correspond to the fitted prediction shifted due to the
  uncertainty on the fluxes \cite{Mention:2011rk}, as indicated in the right
  panel. The dashed lines show the best fit assuming a free overall
  normalization of reactor fluxes. See text for details. \label{fig:SBLreact}}
\end{figure}

The SBL reactor data is summarized in Fig.~\ref{fig:SBLreact}. We show
the observed rate relative to the predicted rate based on old and new
flux calculations. Due to the slightly higher fluxes according to
\cite{Mueller:2011nm} all experiments observe a smaller ratio with the
new fluxes. In Fig.~\ref{fig:SBLreact} we show also the result of a
fit to the data with the predicted fluxes, allowing the four neutrino
fluxes to float in the fit subject to the uncertainties as described
above. In the fit we assume that the experimental systematic errors
of the three data points from Bugey3, G\"osgen, and Krasnoyarsk, as
well as Bugey4 and ROVNO are correlated, due to the same experimental
technique. We obtain $\chi^2 = 8.1 (13.0)$ for 12 degrees of freedom
using old (new) fluxes. Clearly old fluxes provide a better fit to the
data, whereas the $\chi^2$ for new fluxes is still acceptable (P-value
of 37\%). Such a good fit can be obtained by a rescaling of the fluxes
(subject to the quoted uncertainties) as shown in the right panel.

The dashed lines in the figure correspond to a fit where we introduce an
overall factor $f$ in front of the fluxes, which we let float freely in the
fit. For the old fluxes we find the best fit value of $f = 0.984$ with
$f=1$ within the 1$\sigma$ range. In contrast, for new fluxes we obtain $f =
0.942 \pm 0.024$, and $f=1$ disfavored with $\Delta\chi^2 = 6.2$ which
corresponds to about $2.5\sigma$. This is the origin
of the ``reactor anti-neutrino anomaly'' discussed in \cite{Mention:2011rk}.
A possible explanation of this anomaly could be the presence of a
sterile neutrino ``visible'' in this oscillation channel but not in
the solar and/or atmospheric conversions, the so-called 3+1 scenario,
see e.g.~\cite{Maltoni:2001mt}. However, within the uncertainties on
the neutrino flux prediction, the goodness of fit of the new fluxes to
SBL reactor data is still rather good. Given this somewhat ambiguous
situation, in the following we will present results for 3-flavour
oscillations adopting different assumptions on reactor neutrino
fluxes: $(a)$ motivated by the excellent goodness of fit of SBL data
to the new flux prediction we take fluxes and the quoted uncertainties
at face value, and $(b)$ we introduce the free flux normalization $f$
in the fit. This second option takes into account the possible
presence of a sterile neutrino or some other correlated effect on all
reactor neutrino fluxes. In this scenario the SBL reactor experiments
effectively serve as near detectors determining the flux which is then
used as input for the oscillation analysis at longer baselines.

\begin{figure}
  \centering
  \includegraphics[height=6cm]{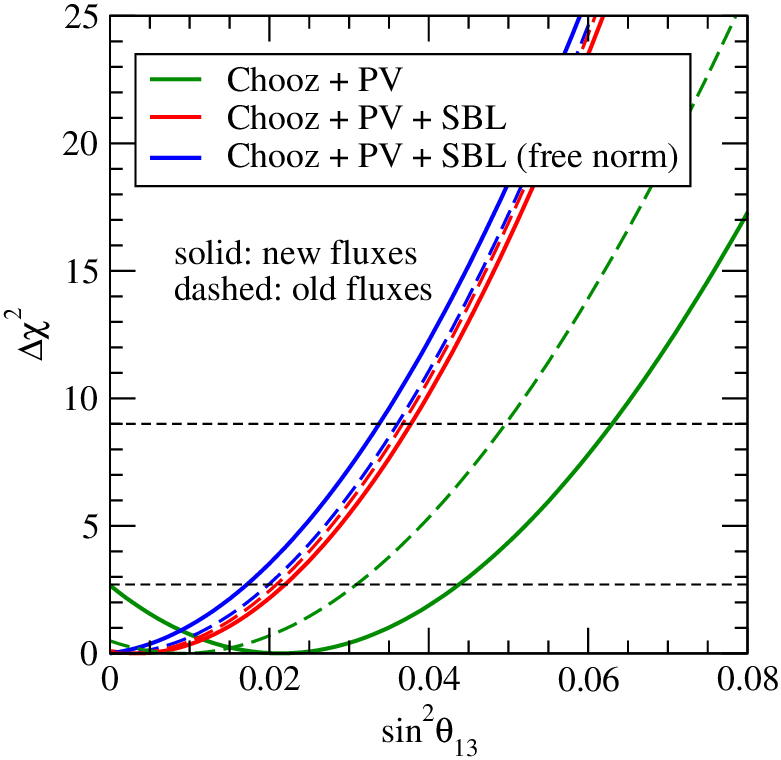}\qquad
  \includegraphics[height=6cm]{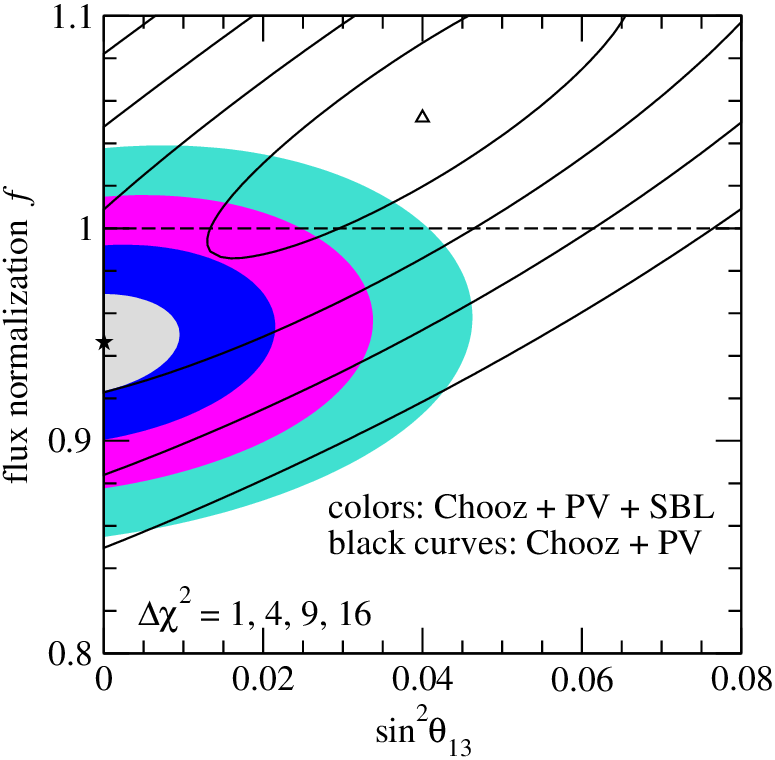}
  \caption{Left: $\Delta\chi^2$ as a function of $\sin^2\theta_{13}$
    for the Chooz and Palo Verde (PV) reactor experiments (green), and
    in combination with the short-baseline (SBL) experiments from
    Fig.~\ref{fig:SBLreact}. Solid (dashed) curves refer to the new
    (old) reactor anti-neutrino fluxes. For the blue curves (``free
    norm'') a free overall normalization factor has been introduced
    for the reactor fluxes. In this figure we fix $\Delta m^2_{31} =
    2.45\times 10^{-3}$~eV$^2$. Right: Contours in the plane of
    $\sin^2\theta_{13}$ and the flux normalization $f$. Colored
    regions (curves) correspond to Chooz + PV with (without) including
    the SBL experiments. \label{fig:th13-react}}
\end{figure}

We show the $\Delta\chi^2$ from the Chooz and Palo Verde experiments
as a function of $\sin^2\theta_{13}$ in Fig.~\ref{fig:th13-react}
(left) for various assumptions on the fluxes. If the new fluxes are
taken at face value and SBL reactor experiments are not included in
the fit (solid green curve), we obtain from Chooz and Palo Verde a
hint for $\theta_{13} > 0$ at about 90\%~CL, with a best fit
  value at $\sin^2\theta_{13} = 0.021$. In this case $\bar\nu_e$
disappearance due to $\theta_{13}$ accounts for the suppression of the
observed rate at $L\simeq 1$~km relative to the slightly increased
prediction from the new fluxes. However, as soon as SBL reactor
experiments are included in the fit, the hint essentially disappears
and $\theta_{13} = 0$ is consistent within 1$\sigma$. This can be
understood from Fig.~\ref{fig:SBLreact}, which shows that the SBL
reactor experiments pull down the flux predictions, leaving less room
for a suppression at 1~km due to $\theta_{13}$. The upper limit on
$\sin^2\theta_{13}$ is very similar with old and new fluxes
  when SBL data are included, irrespective of whether the
  normalization is left free or not. These results are in agreement
  with Ref.~\cite{Mention:2011rk}. Fig.~\ref{fig:th13-react} (right)
shows the correlation between $\sin^2\theta_{13}$ and the flux
normalization $f$ with and without SBL experiments.

To summarize this section we emphasize that due to the tension between
the new flux predictions and the short baseline oscillation data,
including or not the SBL reactors in the fit leads to different
results concerning the extracted value of $\theta_{13}$. This will
indeed be seen in our subsequent fit results, see for example,
Sec.~\ref{sec:global}.

\section{Solar + KamLAND analysis in the light of new reactor fluxes}
\label{sec:solar}

\begin{figure}
  \centering
  \includegraphics[height=6cm]{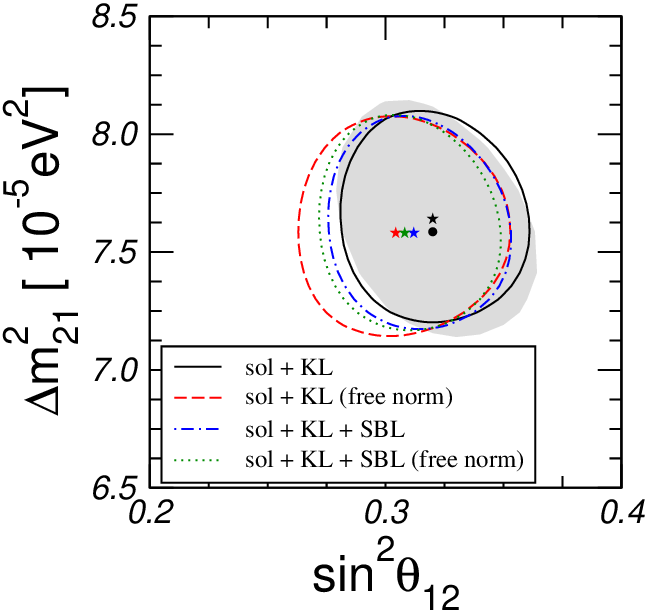}
  \includegraphics[height=6cm]{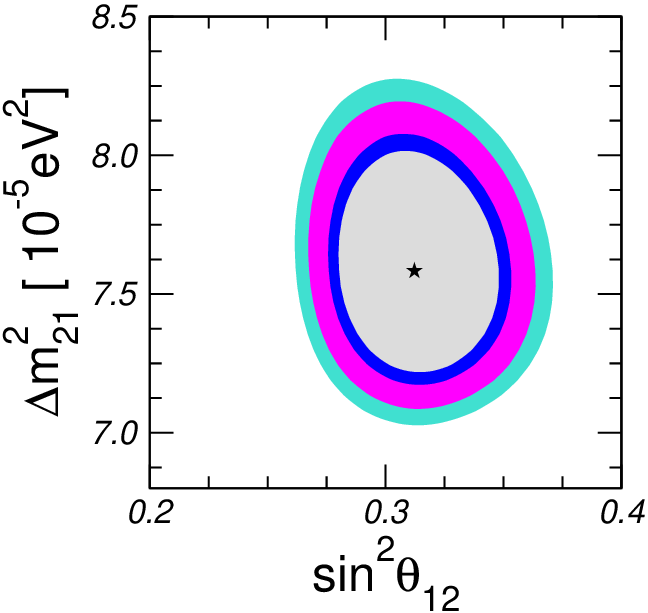}
  \caption{Left: 2$\sigma$ allowed regions in the $\sin^2\theta_{12} -
  \Delta m^2_{21}$ plane from the analysis of solar + KamLAND data
  minimizing over $\theta_{13}$. The different curves show the results
  obtained using different assumptions for reactor data, as indicated in the
  legend. For comparison we show as grey-shaded area the region obtained in
  our previous solar + KamLAND data analysis.  Right: region allowed at 
  90\%, 95\%, 99\%, 99.73\%~CL in our recommended analysis of solar +
  KamLAND data including SBL reactor results. \label{fig:s12-dm21}}
\end{figure}

\begin{figure}
  \centering
  \includegraphics[height=6cm]{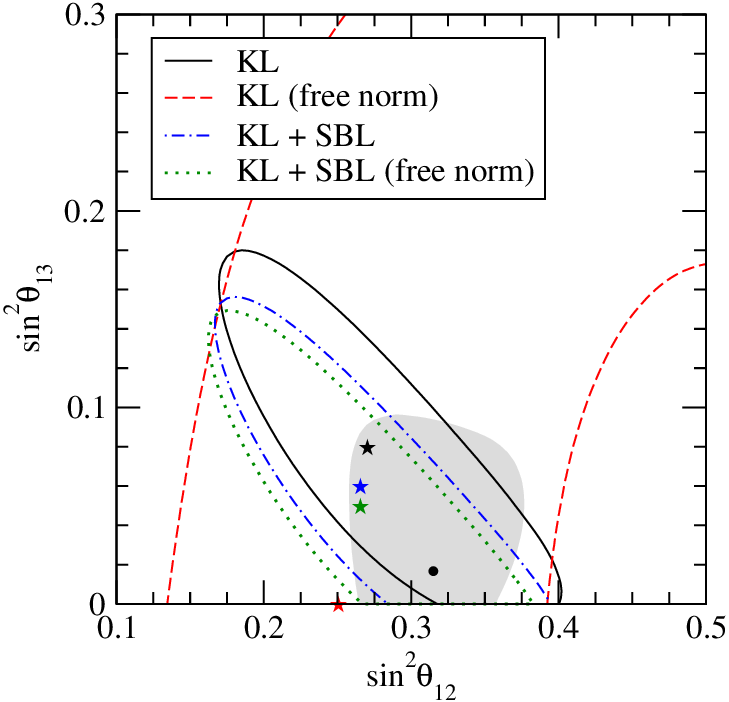}
  \includegraphics[height=6cm]{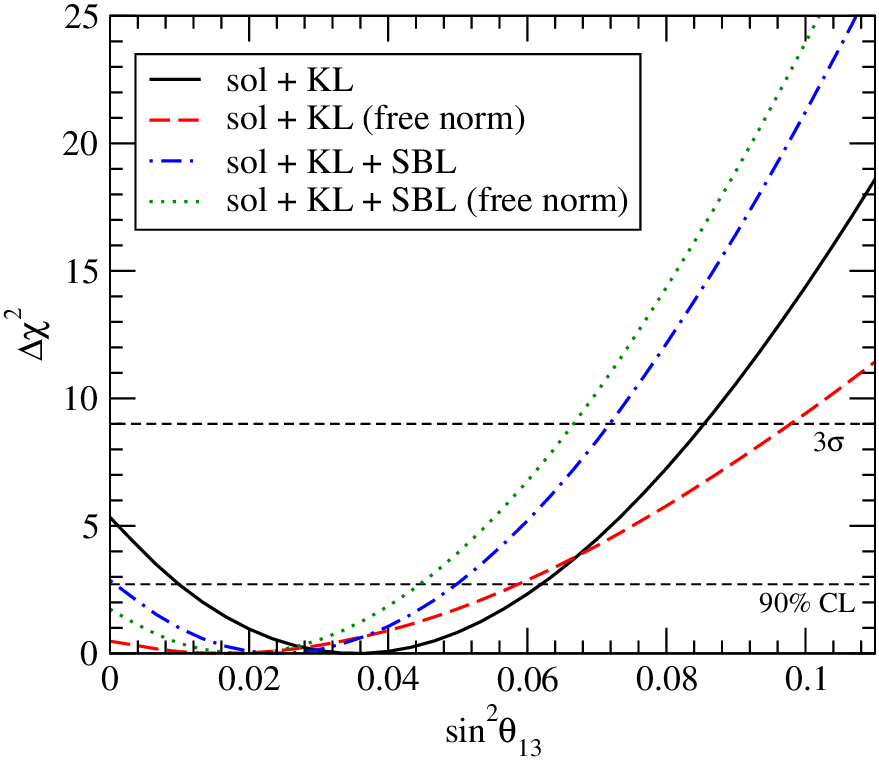}
  \caption{Left: 2$\sigma$ allowed region in the $\sin^2\theta_{12}-\sin^2\theta_{13}$ plane
    obtained from KamLAND data using different assumptions on the reactor data
    analysis. For comparison we also show the 2~$\sigma$ allowed region from
    solar data only (grey-shaded area). Right: $\Delta\chi^2$ as a function of
    $\sin^2\theta_{13}$ for the solar + KamLAND data analysis under the same 
    assumptions as in the left panel.
    \label{fig:s13-sol+kl}}
\end{figure}

The release of new solar data from the second and third phase of the
Super-Kamiokande~\cite{:2008zn,Abe:2010hy} experiment, new data from
the reactor experiment KamLAND~\cite{Gando:2010aa}, and the new
predictions for the reactor anti-neutrino fluxes require a full
revision of the solar + KamLAND neutrino data analysis.
In Fig.~\ref{fig:s12-dm21} we show the 2$\sigma$ allowed region from
the solar + KamLAND neutrino data using different assumptions in the
reactor data analysis, such as including or not including the
short-baseline data or the use of a free normalization factor for the
reactor anti-neutrino fluxes. In all cases the shift of the allowed
region is not very significant, and the best fit point values vary
between 0.304 and 0.320 for $\sin^2\theta_{12}$ while the solar mass
splitting $\Delta m^2_{21}$ goes from 7.59 to $7.64\times10^{-5}$~eV$^2$.
The variation in $\sin^2\theta_{12}$  is a bit larger because this
parameter is correlated with the shift on  $\sin^2\theta_{13}$, as
shown in Fig.~\ref{fig:s13-sol+kl}.

In the left panel of Fig.~\ref{fig:s13-sol+kl} we show the 2$\sigma$ allowed
regions in the $\sin^2\theta_{12}-\sin^2\theta_{13}$ plane for different
choices of the reactor data analysis. For comparison we also show in grey
the 2$\sigma$ allowed region obtained from solar data only. In the right
panel of Fig.~\ref{fig:s13-sol+kl} we show the constraints on $\theta_{13}$
from the combination of solar and KamLAND data. For the global neutrino
oscillation fit presented in the following we will use the analysis of
KamLAND + SBL data without free normalization, labelled as ``sol+KL+SBL'' in
Fig.~\ref{fig:s13-sol+kl}. In this case we get the following best fit value
for $\theta_{13}$:
 \begin{equation}
\sin^2\theta_{13} = 0.023^{+0.016}_{-0.013} \quad \text{(solar + KamLAND)}
\end{equation}
with $\Delta\chi^2 (\sin^2\theta_{13} = 0)$ = 2.9, and therefore a
1.7$\sigma$ hint for $\theta_{13} \neq 0$ coming from the solar sector.
Comparing with our previous analysis~\cite{Schwetz:2008er}, the inclusion of
new solar and KamLAND data and the new reactor fluxes results in a
similar best fit value for the $\theta_{13}$ mixing angle (before we got
$\sin^2\theta_{13} = 0.022$), but a slightly larger significance for
non-zero $\theta_{13}$ (before: $\Delta\chi^2$= 2.2). The origin for
this is mainly the preference of KamLAND data for a non-zero $\theta_{13}$
visible in the left panel of Fig.~\ref{fig:s13-sol+kl}. For KamLAND
$\theta_{13}$ acts mainly as an overall reduction of the spectrum, and
therefore the increased event rate due to the new reactor fluxes can be
compensated by a non-zero $\theta_{13}$ in KamLAND. 

Let us mention that Ref.~\cite{Mention:2011rk} also investigates the
implication for $\theta_{13}$ of reactor neutrino data in the light of the
new fluxes. Our results are in reasonable agreement, although minor
quantitative differences occur presumably due to different data used as well
as different analysis strategies.

\section{Global 3-neutrino analysis and status of $\theta_{13}$ }
\label{sec:global}

Let us now present the results of the global analysis combining all data
mentioned in the previous sections~\footnote{ See also
Refs.~\cite{Schwetz:2008er,Maltoni:2004ei} for previous experimental
references.}. As default for the reactor analysis we use the new
anti-neutrino flux predictions and include in the analysis the SBL reactor
experiments as discussed in Sec.~\ref{sec:reactor}.

\begin{figure}
  \centering
  \includegraphics[height=5.5cm]{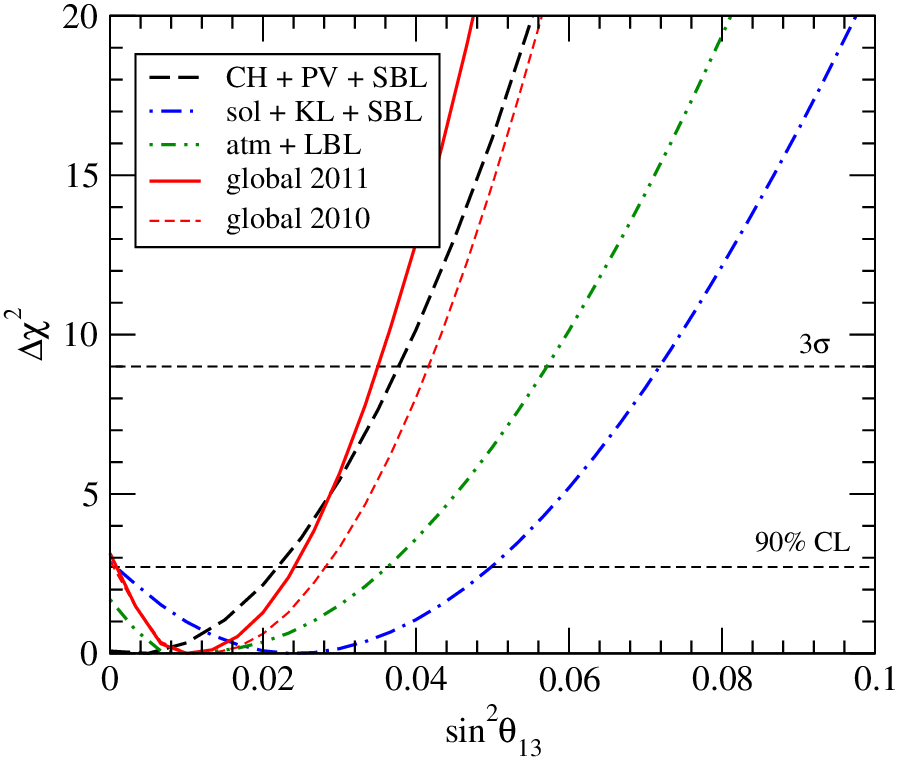}
  \includegraphics[height=5.5cm]{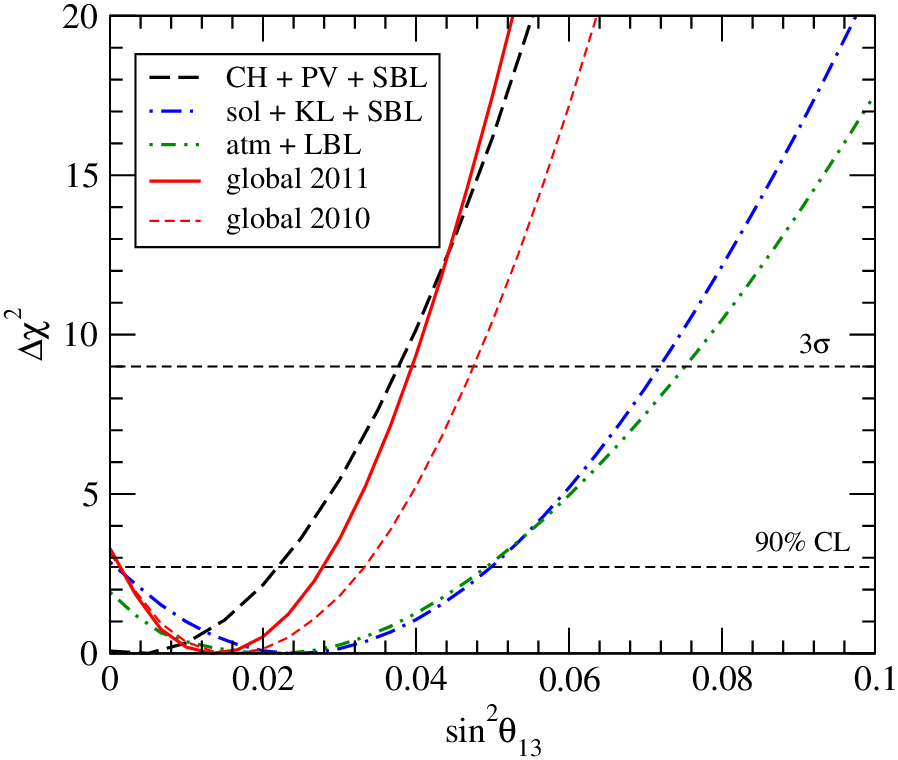}
  \caption{Constraint on $\sin^2\theta_{13}$ from different data
    sets, shown for NH (left) and IH (right). The curves labeled
    ``CH+PV+SBL'' include the Chooz, Palo Verde and the short-baseline reactor
    experiments,  ``solar+KL+SBL''  include solar, KamLAND and short-baseline
    reactor data, and ``atm + LBL'' include Super-K atmospheric data,
    MINOS (disappearance and appearance), and K2K. The results from
    our previous 2010 analysis are also shown for comparison. 
\label{fig:th13-sum11-tot}}
\end{figure}

Fig.~\ref{fig:th13-sum11-tot} shows the $\chi^2$ profile as a function
of $\sin^2\theta_{13}$ for various data samples. In the upper part of
Table~\ref{tab:sq13} we display the corresponding best fit values and
the significance for $\theta_{13} > 0$. In our standard recommended
analysis (new reactor fluxes, SBL reactors included) we find no
significant hint from Chooz and Palo Verde data, but the $1.7\sigma$
hint from solar + KamLAND (see Sec.~\ref{sec:solar}) and the
$1.3\sigma$ ($1.4\sigma$ for IH) hint from atmospheric + MINOS data
(see Sec.~\ref{sec:atm-th13}) combine to a global hint at $1.8\sigma$
for NH and IH, to be compared with the $1.5\sigma$ obtained in our
previous analysis. In the lower part of Table~\ref{tab:sq13} we
discuss how this result depends on details of the reactor neutrino
analysis. If SBL data are not used in the fit the significance for
$\theta_{13}>0$ is pushed close to $3\sigma$ because as discussed in
Sec.~\ref{sec:reactor} in this case also Chooz and Palo Verde prefer
$\theta_{13} > 0$ at about 90\%~CL. However, in the flux-free reactor
analysis as well as with the old reactor fluxes the hint decreases to
about $1.4\sigma$ and $1.8\sigma$, respectively. The entry in the
table labeled ``global without reactors'' comes from atmospheric and
solar neutrinos plus data from the MINOS long-baseline
experiment. Therefore, these results are independent of any ambiguity
due to reactor fluxes, and we observe that a non-trivial limit on
$\theta_{13}$ emerges even in this case. Let us mention that here we
always assume the AGSS09 solar model~\cite{Serenelli:2009yc}. As
discussed previously~\cite{Schwetz:2008er, GonzalezGarcia:2010er}
there is a minor dependence of the hint for $\theta_{13}$ on this
assumption.

\begin{table}[ht]\centering
    \catcode`?=\active \def?{\hphantom{0}}
    \begin{tabular}{|>{\rule[-2mm]{0pt}{6mm}}l|c|c|c|}
        \hline
& $\sin^2\theta_{13}$ &  $\Delta\chi^2(\theta_{13}=0)$ &
        3$\sigma$ bound \\
        \hline
        solar + KamLAND + SBL & $0.023^{+0.016}_{-0.013}$ & 2.9
        (1.7$\sigma$) & 0.072\\
        Chooz + Palo Verde + SBL & $0.005^{+0.010}_{-0.020}$ & 0.07
        (0.26$\sigma$) & 0.038\\
    \hline
        atmospheric + MINOS &
        \begin{tabular}{c}
      $0.010^{+0.016}_{-0.008}$\\
         $0.020^{+0.018}_{-0.015}$
    \end{tabular}
    &
    \begin{tabular}{c}
      1.7 (1.3$\sigma$)\\
      1.9 (1.4$\sigma$)
    \end{tabular}
    &
    \begin{tabular}{c}
      0.057\\
      0.075
    \end{tabular}\\
\hline
    global without reactors &
    \begin{tabular}{c}
      $0.013^{+0.014}_{-0.009}$\\
      $0.020^{+0.015}_{-0.012}$
    \end{tabular}
    &
    \begin{tabular}{c}
      2.3 (1.5$\sigma$)\\
      2.7 (1.6$\sigma$)
    \end{tabular}
    &
    \begin{tabular}{c}
         0.053\\
         0.065
       \end{tabular}\\
        \hline
        global with SBL &
        \begin{tabular}{c}
      $0.010^{+0.009}_{-0.006}$ \\
          $0.013^{+0.009}_{-0.007}$
        \end{tabular}
&
        \begin{tabular}{c}
      3.1 (1.8$\sigma$) \\
          3.3 (1.8$\sigma$)
        \end{tabular}
&
        \begin{tabular}{c}
      0.035\\
          0.039
        \end{tabular}\\
        \hline
        global with SBL (free norm) &
        \begin{tabular}{c}
          $0.007^{+0.009}_{-0.005}$ \\
          $0.010^{+0.009}_{-0.007}$
        \end{tabular}
&
        \begin{tabular}{c}
          2.0 (1.4$\sigma$) \\
          1.9 (1.4$\sigma$)
        \end{tabular}
&
        \begin{tabular}{c}
          0.032\\
          0.037
        \end{tabular}\\
        \hline
        global without SBL &
        \begin{tabular}{c}
          $0.020^{+0.010}_{-0.008}$\\
          $0.027^{+0.009}_{-0.010}$
        \end{tabular}
&
        \begin{tabular}{c}
          7.0 (2.6$\sigma$)\\
          8.0 (2.8$\sigma$)
        \end{tabular}
&
        \begin{tabular}{c}
          0.048 \\
          0.054
        \end{tabular}\\
        \hline
        global without SBL (old fluxes) &
        \begin{tabular}{c}
          $0.012^{+0.010}_{-0.007}$\\
          $0.017\pm0.010$
        \end{tabular}
&
        \begin{tabular}{c}
       2.9 (1.7$\sigma$)\\
           3.2 (1.8$\sigma$)
        \end{tabular}
&
        \begin{tabular}{c}
       0.042\\
           0.048
        \end{tabular}\\
        \hline
\end{tabular}  
\caption{\label{tab:sq13} The best fit values for $\sin^2\theta_{13}$ with
  $1\sigma$ errors, the significance of the $\theta_{13} >
  0$  hint, and the upper bound on $\sin^2\theta_{13}$ at $3\sigma$ for different
  data samples and for different reactor neutrino data assumptions. For a given global analysis the upper
  (lower) numbers refer to normal (inverted) neutrino mass hierarchy. We
  always use the new reactor fluxes \cite{Mueller:2011nm} except for the row
  labeled ``old fluxes'' which uses previous results \cite{schreckenbach}.
  The row labeled ``free norm'' assumes a free reactor
  anti-neutrino flux normalization.  }
\end{table}

\begin{figure}
  \centering 
  \includegraphics[height=6cm]{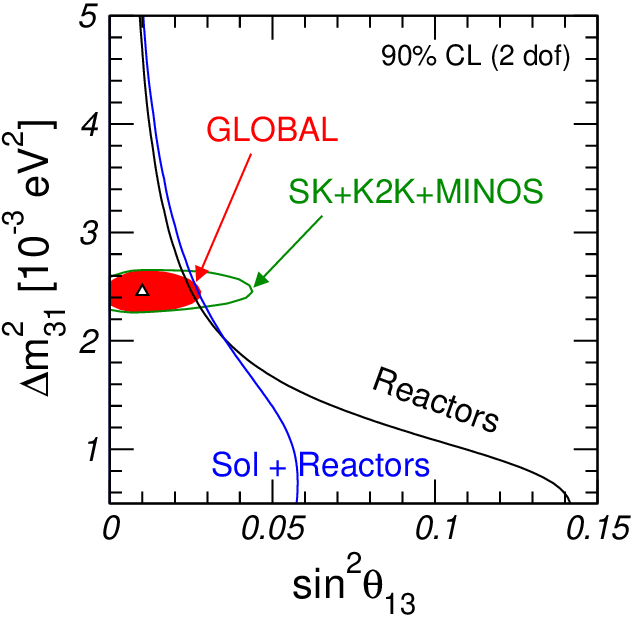}
  \includegraphics[height=6cm]{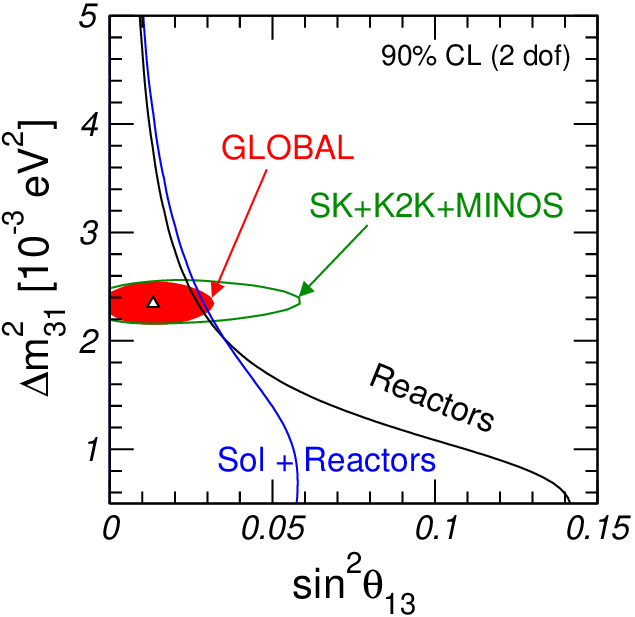} 
  \caption{Illustration of the interplay of the global data on the
  $\sin^2\theta_{13}$ bound. Left: NH and right: IH.
  \label{fig:th13-contour} }
\end{figure}

\begin{table}[ht]\centering
    \catcode`?=\active \def?{\hphantom{0}}    
    \begin{tabular}{|@{\quad}>{\rule[-2mm]{0pt}{6mm}}l@{\quad}|@{\quad}c@{\quad}|@{\quad}c@{\quad}|@{\quad}c@{\quad}|}
        \hline
        parameter & best fit $\pm 1\sigma$ & 2$\sigma$ & 3$\sigma$ 
        \\
        \hline
        $\Delta m^2_{21}\: [10^{-5}\eVq]$
        & $7.59^{+0.20}_{-0.18}$  & 7.24--7.99 & 7.09--8.19 \\[3mm] 
        $\Delta m^2_{31}\: [10^{-3}\eVq]$
        &
	\begin{tabular}{c}
	  $2.45\pm0.09$\\
	  $-(2.34^{+0.10}_{-0.09})$
	\end{tabular}
	& 
	\begin{tabular}{c}
	  $2.28-2.64$\\
	  $-(2.17-2.54)$
	\end{tabular}
	& 
	\begin{tabular}{c}
	  $2.18-2.73$\\
	  $-(2.08-2.64)$
	\end{tabular}
	\\[6mm] 
        $\sin^2\theta_{12}$
        & $0.312^{+0.017}_{-0.015}$ & 0.28--0.35 & 0.27--0.36\\[3mm]  
        $\sin^2\theta_{23}$	
        & 
	\begin{tabular}{c}	
	  $0.51\pm0.06$\\
	  $0.52\pm0.06$
	\end{tabular}	  
	& 
	\begin{tabular}{c}
          0.41--0.61\\
	  0.42--0.61
        \end{tabular}
	& 0.39--0.64 \\[5mm] 
        $\sin^2\theta_{13}$
        &
	\begin{tabular}{c}
	  $0.010^{+0.009}_{-0.006}$\\
	  $0.013^{+0.009}_{-0.007}$
	\end{tabular}  
	& 
	\begin{tabular}{c}
	$\leq$ 0.027\\
	$\leq$ 0.031
	\end{tabular}  
	& 
	\begin{tabular}{c}
	$\leq$ 0.035\\ 
	$\leq$ 0.039
	\end{tabular}\\
        \hline
\end{tabular}
  \caption{ \label{tab:summary2011} Neutrino oscillation parameters
    summary. For $\Delta m^2_{31}$, $\sin^2\theta_{23}$, and
  $\sin^2\theta_{13}$ the upper (lower) row corresponds to normal (inverted)
  neutrino mass hierarchy. We assume the new reactor anti-neutrino fluxes
  \cite{Mueller:2011nm} and include short-baseline reactor neutrino experiments in the fit.}
\end{table}

Fig.~\ref{fig:th13-contour} illustrates the interplay of the various
data sets in the plane of $\sin^2\theta_{13}$ and $\Delta
m^2_{31}$. In Table~\ref{tab:summary2011} we summarize the
determination of neutrino oscillation parameters for our reference
default reactor analysis. As discussed in Sec.~\ref{sec:solar} the
impact of this choice upon the leading oscillation parameters is
small. We find that inverted hierarchy gives a slightly better fit,
however, with only $\Delta\chi^2 = 0.54$ with respect to the best fit
in normal hierarchy.

\section{Summary and conclusions}
\label{sec:summary-conclusions}

We have presented an updated global fit to world's neutrino oscillation
data.  The recent re-evaluation of the anti-neutrino fluxes emitted in
nuclear power plants \cite{Mueller:2011nm} introduces some ambiguity in the
results obtained for the mixing angle $\theta_{13}$. Since the new
predictions are in slight disagreement with data from short-baseline (SBL)
reactor experiments, with $L \lesssim 100$~m, it becomes necessary to
include these data in the fit. A flux-free analysis of SBL data prefers an
off-set of the reactor neutrino flux of about 6\% from the predicted value
with a significance of about $2.5\sigma$. Taken at face value this might
indicate either some un-accounted systematic effect (either in the new
calculations or in the reactor data), or even the presence of some kind of
new physics such as sterile neutrino oscillations with $\Delta m^2 \sim
1$~eV$^2$~\cite{Mention:2011rk}. Here we stick to the three-flavour
framework, the sterile neutrino hypothesis will be discussed
elsewhere~\cite{Kopp:2011qd}.

Despite this hint for a deviation of the observed reactor
anti-neutrino flux from its prediction, the goodness-of-fit of the SBL
reactor neutrino data with the new fluxes is still very good
($\chi^2=13$ for 12 degrees of freedom). Motivated by this result, we
adopt as our recommended default analysis the new fluxes and include
the SBL data in the fit. In such a way we obtain a hint for
$\theta_{13} > 0$ at $1.8\sigma$, coming from a preference for a
finite $\theta_{13}$ from KamLAND data combined with a somewhat weaker
hint from the joint analysis of atmospheric + MINOS data. In
Table~\ref{tab:sq13} we show in detail how the global result depends
on the assumptions on the reactor neutrino analysis, yielding hints
for $\theta_{13} > 0$ ranging between 1.4$\sigma$ and 2.8$\sigma$,
with best fit values between $\sin^2\theta_{13} = 0.007$ and
$0.027$. This somewhat ambiguous situation regarding $\theta_{13}$
emerges from the slight tension due to the new reactor flux
predictions with data. It will be interesting to see how the upcoming
results from new reactor as well as accelerator experiments searching
for $\theta_{13}$ will contribute to resolving the issue, see
Ref.~\cite{Mezzetto:2010zi} for an overview and references.

The main results of our recommended default analysis of three-neutrino
oscillation parameters are summarized in Table.~\ref{tab:summary2011}
and in the right panels of Figs.~\ref{fig:atm-sum11} and
\ref{fig:s12-dm21} for the leading ``atmospheric'' and ``solar''
oscillation parameters, as well as in Fig.~\ref{fig:th13-contour} for
the mixing angle $\theta_{13}$.

\section*{Acknowledgments}

We thank Ed Kearns for discussions during Neutrino 2010 and, in particular,
for providing us the Super-Kamiokande Collaboration $\chi^2$ map obtained in
Ref.~\cite{Wendell:2010md}. We also thank T.~Lasserre, M.~Fechner and
D.~Lhuillier for communication on the SBL reactor analysis in relation to
the new fluxes.  Work supported by Spanish grants FPA2008-00319/FPA,
MULTIDARK Consolider CSD2009-00064, PROMETEO/2009/091, and by EU network
UNILHC, PITN-GA-2009-237920. M.T.\ acknowledges financial support from CSIC
under the JAE-Doc programme. This work was partly supported by the
Transregio Sonderforschungsbereich TR27 ``Neutrinos and Beyond'' der
Deutschen Forschungsgemeinschaft.


\bibliographystyle{unsrt}

\end{document}